\documentclass[12pt]{article}
\usepackage{amssymb,amsmath}
\usepackage{graphicx,psfrag}

\newcommand{\bea}{\begin{eqnarray}}
\newcommand{\eea}{\end{eqnarray}}
\newcommand{\simgt}{\hbox{ \raise3pt\hbox to 0pt{$>$}\raise-3pt\hbox{$\sim$} }}
\newcommand{\simlt}{\hbox{ \raise3pt\hbox to 0pt{$<$}\raise-3pt\hbox{$\sim$} }}

\newcommand{\clfn}{\setcounter{footnote}{0}}

\begin{document}

\begin{titlepage}
\title{\bf 
\Large
Family Gauge Symmetry as an Origin of\\ 
Koide's Mass Formula and
Charged Lepton Spectrum\footnote{
Based on the talk given at Karlsruhe University, Germany, in Feb.\ 2009.
}
\vspace{7mm}}
\author{
Y.~Sumino
\\ \\ \\ Department of Physics, Tohoku University\\
Sendai, 980--8578 Japan
}
\date{}
\maketitle
\thispagestyle{empty}
\vspace{-3.8truein}
\begin{flushright}
{\bf TU--842}\\
{\bf Mar. 2009}
\end{flushright}
\vspace{4truein}
\begin{abstract}
{\small
Recently we have proposed mechanisms to explain
origins of the charged lepton spectrum
as well as Koide's mass formula, on the basis of $U(3)\times O(3)$
family gauge symmetry.
In this note, we review the basic ideas of
these mechanisms.
Without technical details, and adding some
speculations, we 
give a sketch of the mechanisms,
what the important points are and what 
assumptions are involved.

We adopt a known scenario, in which
the charged lepton spectrum is determined by
the vacuum expectation value of a 
scalar field that takes values on 3-by-3 matrix.
Within this scenario, we propose a mechanism, in which
the radiative correction induced by $U(3)$
family gauge interaction
cancels the QED
radiative correction to Koide's mass formula.
We consider $SU(9)\times U(1)$ symmetry broken down to
$U(3)\times O(3)$ symmetry.
This leads to a potential model which predicts
Koide's mass formula and the charged lepton spectrum
consistent with the experimental values, by largely avoiding
fine tuning of parameters.
These are discussed within an effective theory, and we 
argue for its validity and usefulness.
}
\end{abstract}
\vfil

\end{titlepage}

\section{Introduction}
\label{s1}
\clfn

In this note, we review the basic ideas presented in our recent works
\cite{Sumino:2008hu,Sumino:2008hy},
which analyze possible connections between the charged lepton
spectrum and family gauge symmetries;
in particular, full advantage of Koide's 
mass formula has been taken to study this connection.

\begin{figure}[h]\centering
\psfrag{111}{$(1,1,1)$}
\psfrag{45}{$\theta=45^\circ$}
\psfrag{rootm}{\small $(\sqrt{m_e},\sqrt{m_\mu},\sqrt{m_\tau})$} 
\includegraphics[width=6cm]{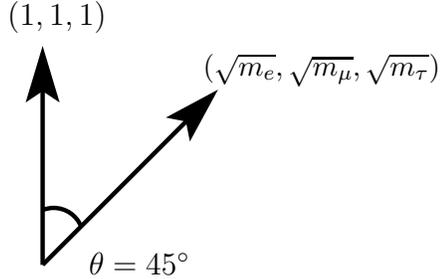}
\caption{\small 
Geometrical interpretation 
of Koide's mass formula, eq.~(\ref{KoideMF}). 
\label{fig1}}
\end{figure}

Koide's mass formula \cite{Koide:1982wm}, found by Koide
in 1982, is an empirical relation
among the charged lepton masses,
which holds with a striking precision.
The formula can be described in the following way:
Consider two vectors $(1,1,1)$
and $(\sqrt{m_e},\sqrt{m_\mu},\sqrt{m_\tau})$ 
in a 3-dimensional space;
then, the angle 
between
these two vectors is equal to $45^\circ$ \cite{Foot:1994yn},
see Fig.~\ref{fig1}\\
\\
Equivalently, the formula is expressed as
\bea
\frac{\sqrt{m_e}+\sqrt{m_\mu}+\sqrt{m_\tau}}{
\sqrt{3\,(m_e+m_\mu+m_\tau)}}
=\cos 45^\circ
=\frac{1}{\sqrt{2}} \, .
\label{KoideMF}
\eea
Present experimental values of the on-shell (pole) masses of the
charged leptons read \cite{Amsler:2008zz}
\bea
&&
m_e=0.510998910\pm 0.000000013~{\rm MeV} \, ,
\\
&&
m_\mu=105.658367\pm 0.000004~{\rm MeV} \, ,
\\
&&
m_\tau = 1776.84 \pm 0.17~{\rm MeV} \, .
\eea
It may be noteworthy that
the accuracy of the tau mass measurement
(which limits the experimental accuracy of Koide's formula) 
is still improving in
the last few years.
Using these values, one finds that
\bea
\sqrt{2}\,
\left[
\frac{\sqrt{m_e}+\sqrt{m_\mu}+\sqrt{m_\tau}}{
\sqrt{3\,(m_e+m_\mu+m_\tau)}}
\right]
=1.000005 \pm 0.000007 \, .
\eea
Thus, Koide's formula is valid within the
current experimental accuracy of 
$7\times 10^{-6}\,$!
We emphasize that
it is the pole masses  that satisfy Koide's formula with this precision.

Given the remarkable accuracy with which Koide's mass formula holds,
many speculations have been raised as to existence of some
physical origin behind
this mass formula 
\cite{Koide:1983qe,Foot:1994yn,
Koide:1995xk,Koide:2005nv,Li:2006et,Xing:2006vk,Ma:2006ht,Rosen:2007rt}.
Despite these attempts,
so far no realistic model or mechanism has been found
which predicts Koide's mass formula within the required accuracy.
A most serious problem one
faces, when speculating physics underlying Koide's formula,
is caused by the QED radiative correction \cite{Xing:2006vk}.
One expects that some physics at a short-distance scale beyond our
current reach
determines the spectrum of the Standard-Model (SM) fermions.
Then it seems more natural that
the relation (\ref{KoideMF}) is satisfied by the running masses
$\bar{m}_i(\mu)$ 
(or the corresponding Yukawa couplings $\bar{y}_i(\mu)$)
renormalized  at a high energy scale $\mu\gg M_W$ than
by the pole masses.
If this is the case, however, the QED radiative correction
violates the relation between the pole masses.
\begin{figure}[h]\centering
\psfrag{gamma}{$\gamma$}
\psfrag{l}{$\ell$}
\includegraphics[width=6cm]{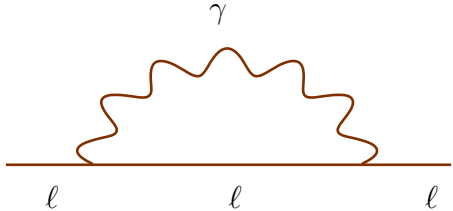}
\caption{\small 
QED 1-loop diagram contributing to the pole mass.
\label{QED1Lradcorr}}
\end{figure}\\
In fact, the 1-loop
QED radiative correction is given by
\bea
m^{\rm pole}_i = \left[
1+\frac{\alpha}{\pi}\left\{
\frac{3}{4}\log\left(
\frac{\mu^2}{\bar{m}_i(\mu)^2}
\right) +1
\right\}
\right]\,
\bar{m}_i(\mu) \, .
\label{QED1Lcorr}
\eea
Here, $\bar{m}(\mu)$ and $m^{\rm pole}$ denote the 
running mass defined in the modified--minimal--subtraction scheme
($\overline{\rm MS}$ scheme) and the pole mass, respectively;
$\mu$ represents the
renormalization scale.
Suppose  $\bar{m}_i(\mu)$ 
satisfy the relation
(\ref{KoideMF}) at a high energy scale $\mu\gg M_W$.
Then $m_i^{\rm pole}$
do not satisfy the same relation \cite{Li:2006et,Xing:2006vk}: 
Eq.~(\ref{KoideMF})
is corrected by approximately 0.1\%, which is 120 times 
larger than the present experimental error.
Note that this correction
originates only from the term $-3\alpha/(4\pi) \times\bar{m}_i \, \log(\bar{m}_i^2)$
of eq.~(\ref{QED1Lcorr}), since the other terms, which are of the form
${\rm const.}\times\bar{m}_i$, do not affect Koide's formula.
This is because, the latter corrections only change the length of the
vector $(\sqrt{m_e},\sqrt{m_\mu},\sqrt{m_\tau})$ but not the direction.
As a result, the QED correction to Koide's mass formula turns out to be 
independent of the UV scale $\mu$.
The $\bar{m}_i \log(\bar{m}_i^2)$ correction
results from the fact that $\bar{m}_i$
plays a role of an infrared (IR) cut--off in the
loop integral.
Hence, the QED correction to Koide's formula stems from this IR region.

The 1--loop weak correction is of the form
${\rm const.}\times\bar{m}_i$ in the leading order of
$\bar{m}_i^2/M_W^2$ expansion;
the leading non--trivial correction is
${\cal O}(G_F \bar{m}_i^3/\pi)$ whose effect is smaller than the 
current experimental accuracy.
Other radiative corrections within the SM
(due to Higgs and would-be Goldstone bosons) are also  negligible.

Among various existing models which attempt to explain origins
of Koide's mass formula, we find a class of models particularly
attractive \cite{Koide:1989jq,Koide:1995pb}.
These are the models which predict the mass matrix of 
the charged leptons to be
proportional to the square of 
the vacuum expectation value (VEV)
of a 9--component
scalar field (we denote it as $\Phi$) written in 
a 3--by--3 matrix form:
\bea
{\cal M}_\ell \propto \langle \Phi \rangle \langle \Phi \rangle
~~~\mbox{with}~~~
\langle \Phi \rangle
=
\left(\begin{array}{ccc}
v_1(\mu)&0&0\\
0&v_2(\mu)&0\\
0&0&v_3(\mu)
\end{array}\right) 
\, .
\label{MasPhisq}
\eea
Thus, $(\sqrt{m_e},\sqrt{m_\mu},\sqrt{m_\tau})$ is proportional
to the diagonal elements $(v_1,v_2,v_3)$
of $\langle \Phi \rangle$ in the
basis where it is diagonal.
The above form of the 
lepton mass matrix may originate from an effective higher-dimensional
operator 
\bea
{\cal O}=\frac{\kappa(\mu)}{\Lambda^2}\, 
\bar{\psi}_{Li}\, \Phi_{ik}\, \Phi_{kj}\, \varphi \, e_{Rj} \, .
\label{higherdimopO}
\eea
Here, $\psi_{Li}=(\nu_{Li},e_{Li})^T$ denotes the left--handed lepton 
$SU(2)_L$ doublet
of the $i$--th generation;
$e_{Rj}$ denotes the right-handed charged lepton
of the $j$--th generation;
$\varphi$ denotes the Higgs doublet field;
$\Phi$ is a 9--component scalar field
and is singlet under the SM gauge group.
We suppressed all the indices except for the generation (family)
indices $i,j,k=1,2,3$.
(Summation over repeated indices is understood throughout
the paper.)
The dimensionless Wilson coefficient of this operator is
denoted as $\kappa(\mu)$.
Once $\Phi$ acquires a VEV, the operator $\cal O$ will
effectively be rendered to
the Yukawa interactions of the SM; 
after the Higgs field 
also acquires a VEV, $\langle\varphi\rangle=(0,v_{\rm ew}/\sqrt{2})^T$
with $v_{\rm ew}\approx 250$~GeV, the operator will induce the
charged--lepton mass matrix of the form
eq.~(\ref{MasPhisq}) at tree level.
We assume that the dimension-4 Yukawa interactions
$y_{ij}\,\bar{\psi}_{Li}\varphi e_{Rj}$ are fobidden by
some mechanism; this will be imposed
by symmetry in our scenario to be discussed through Secs.~3--5.

\begin{figure}[h]\centering
\includegraphics[width=6cm]{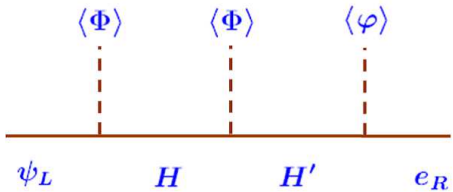}
\caption{\small 
Diagram showing generation of the higher-dimensional operator
${\cal O}=\frac{\kappa(\mu)}{\Lambda^2}\, 
\bar{\psi}_{Li}\, \Phi_{ik}\, \Phi_{kj}\, \varphi \, e_{Rj} $
through see-saw mechanism.
$H$ and $H'$ denote heavy fermions to be integrated out.
\label{see-saw}}
\end{figure}

As an example of underlying theory that leads to
the higher-dimensional operator $\cal O$, we may consider
see-saw mechanism, as depicted in Fig.~\ref{see-saw} 
\cite{Koide:1989jq,Koide:1995xk}.
In this case, the operator $\cal O$ is induced
after 
integrating out the heavy fermions $H$ and $H'$.
\\

On the other hand, the VEV $\langle \Phi \rangle$
is determined by minimizing the potential of scalar fields
in each model.
By deliberately choosing 
a specific form of the potential, the VEV is made to satisfy the relation
\bea
\frac{v_1(\mu)+v_2(\mu)+v_3(\mu)}
{\sqrt{3\,[v_1(\mu)^2+v_2(\mu)^2+v_3(\mu)^2]}}
=\frac{1}{\sqrt{2}}
\label{relvi}
\eea
in the basis where it is diagonal.
Hence, the origin of Koide's formula is
attributed to the specific form of the potential which realizes
this relation in the vacuum configuration.
Up to now, however, no existing model is complete with respect to symmetry.
Namely, every model requires either
absence or strong suppression of
some of the terms in the potential (which are allowed by
the symmetry of that model),
without 
justification.

In our study, we adopt a similar scenario for generating the charged
lepton spectrum.
We introduce a higher-dimensional operator similar to ${\cal O}$
of eq.~(\ref{higherdimopO}) within an effective field theory (EFT)
valid below some cut-off scale $\Lambda(\gg M_W)$.
We analyze a potential of scalar fields within this EFT and
compute the spectrum of the charged leptons.
We compute various radiative corrections and other 
types of corrections
within this EFT.

In the next section (Sec.~2), we explain philosophy of our analysis using
EFT and argue for its validity and usefulness.
In Sec.~3, we explain the mechanism how to cancel the QED correction
in terms of the radiative correction induced by family gauge symmetry.
In Sec.~4, we present a potential model within EFT which leads to
Koide's mass formula and a realistic charged lepton spectrum.
Summary and discussion are given in Sec.~5.

\section{Why EFT? Virtue and assumptions}
\clfn

Let us explain philosophy of our analysis using EFT.
Conventionally a more standard approach for explaining Koide's mass formula has been
to construct models within
renormalizable theories.
In comparison, it 
is certainly a retreat to make an analysis within EFT.
Nevertheless, the long history since the discovery of Koide's formula shows 
that it is quite difficult to construct
a viable renormalizable model for explaining Koide's relation.
It is likely that we are missing some essential hints to achieve this goal, 
if the relation is not a sheer coincidence.
The point we want to make through our study is that within EFT, explanation of Koide's
formula is possible, by largely avoiding fine tuning
of parameters.
Consistency conditions (with respect to symmetries of the theory)
can be satisfied relatively easily in EFT, or in other words, they can be 
replaced by reasonable boundary conditions of EFT at the cut-off scale
$\Lambda$ without conflicting symmetry requirements of the theory.
(See Sec.~4.)
Even under this less restrictive theoretical constraints,
we may learn some important hints concerning the relation between
the lepton spectrum and family symmetries.
These are the role of
specific family gauge symmetry in canceling the QED correction,
the role of family symmetry in stabilizing Koide's mass relation, or
the role of family symmetry in realizing a realistic 
charged lepton spectrum consistently with experimental values.
These properties do not come about separately but 
are closely tied with each other.
These features do not seem to depend on details of more fundamental
theory above the cut-off scale $\Lambda$ but rather on some general
aspects of family symmetries and their breaking patterns.
Thus, we consider that our approach based on EFT would be useful
even in the case
in which physics above the scale $\Lambda$ is obscure and may
involve some totally
unexpected ingredients -----
as it was the case with chiral perturbation theory
before the discovery of QCD.

Before discussing radiative corrections within EFT, one would be worried
about effects of higher-dimensional operators suppressed in higher powers
of $1/\Lambda$.
Indeed, using the values of tau mass and the electroweak symmetry breaking
scale $v_{\rm ew}$, one readily finds that 
$v_3/\Lambda \simgt 0.1$.
Hence, naive dimensional analysis indicates that there would be
corrections to Koide's formula of order 10\% even at tree level.
We now argue that this is not necessarily the case within the scenario
under consideration.
Let us 
divide the corrections into two parts.
These are (i) $1/\Lambda^n$ corrections to the operator
$\cal O$ of eq.~(\ref{higherdimopO}) (those operators which
reduce to the SM Yukawa interactions after $\Phi$ is replaced by
its VEV), and
(ii) $1/\Lambda^n$ corrections to the relation (\ref{relvi})
satisfied by the VEV of $\Phi$.

Concerning the corrections (i), we may consider the following
example.
Suppose that the operator $\cal O$ is induced
from the interactions
\bea
{\cal L}=y_1 \,\bar{\psi}_{Li}\Phi_{ij} H_{Rj} + M\, \bar{H}_{Ri} H_{Li} + 
y_2\, \bar{H}_{Li} \Phi_{ij} H'_{Rj} + M' \bar{H}'_{Ri} H'_{Li} +
y_3\, \bar{H}'_{Li} \varphi e_{Ri}
+ ({\rm h.c.})
\eea
through the diagram shown in Fig.~\ref{see-saw},
after fermions $H_{L,R}$ and $H'_{L,R}$ have been integrated out.
Fermions $H_{L,R}$ and $H'_{L,R}$ are assigned to appropriate
representations of the SM gauge group such that the above interactions
become gauge singlet.
For instance, in the case that 
$v_3/M' \simgt 3$, $y_1,y_2,y_3\approx 1$ and
$v_{\rm ew}/M'<3\times 10^{-3}$, 
one finds,
 by computing the mass eigenvalues,\footnote{
Since the values of $m_\tau$ and $v_{\rm ew}$ are known,
once we choose the values
of $v_3/M'(\simgt 3)$ and $y_1,y_2,y_3(\approx 1)$, the value of
$v_3/M(\simlt 0.03)$ will be fixed.
Then the mass eigenvalues corresponding to 
the SM charged leptons
can be computed in series expansion
in the small parameters
$v_{\rm ew}/M'$, $v_i/M$ and $v_i^2/(MM')=\sqrt{2}m_i/v_{\rm ew}$.
}
that the largest correction to the lepton spectrum 
eq.~(\ref{MasPhisq}) arises from the operator
$\displaystyle
-\frac{y_1^3y_2^3y_3}{2M^3M'^3}\,
\bar{\psi}_L \Phi^6 \varphi e_R
$; its contribution to the
tau mass is
$\delta m_\tau/m_\tau = (m_\tau/v_{\rm ew})^2
\approx 5\times 10^{-5}$.
This translates to a correction to Koide's relation of $3\times 10^{-6}$,
since there is an additional suppression factor due to the fact
$m_e,m_\mu\ll m_\tau$.\footnote{
Note that, in the limit $m_e,m_\mu \to 0$, the direction of 
$(\sqrt{m_e},\sqrt{m_\mu},\sqrt{m_\tau})$ becomes unaffected by 
a correction to
$m_\tau$.
}
Thus, this is an example of underlying mechanism that generates the operator
$\cal O$ without generating higher-dimensional operators conflicting the current
experimental bound.\footnote{
Since in this example $M'$ is not large, it cannot be
regarded as ``see-saw mechanism''.
Nevertheless, we may still construct an EFT in which
the fermions $H_{L,R}$ and $H'_{L,R}$ have been integrated out.
}
If we introduce even more (non--SM) fermions to generate
the leading--order operator $\cal O$, one can always find a pattern of
spectrum of these fermions, for which higher--dimensional operators
are sufficiently suppressed, since the number of 
adjustable parameters increases.
In general, sizes of higher-dimensional operators 
depend heavily on underlying dynamics above the cut-off scale.
(See \cite{Ma:2006ht} for another example of
underlying mechanism.)

\begin{figure}[t]\centering
\psfrag{Phi}{\hspace{0mm}$\Phi$}
\psfrag{psiL}{\hspace{0mm}$\psi_L$}
\psfrag{eR}{\hspace{0mm}$e_R$}
\includegraphics[width=13cm]{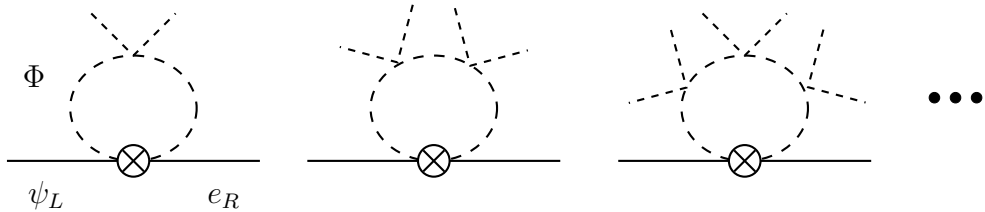}
\caption{\small 
EFT 1-loop diagrams which generate higher-dimensional operators
contributing to the charged lepton spectrum.
Dashed line represents $\Phi$;
$\otimes$ represents
the higher-dimensional
operator which generate the charged
lepton masses at tree-level
[corresponding to ${\cal O}$ of eq.~(\ref{higherdimopO})].
\label{1LoopAnalyInEFT}}
\end{figure}

Let us restrict ourselves within EFT.
If we introduce only the operator $\cal O$, by definition this is the only
contribution to the charged lepton spectrum at tree level.
Whether loop diagrams induce higher-dimensional operators
which violate Koide's relation is an important question,
and a detailed analysis is necessary.
This is the subject of the present study, where
the result depends on the mechanisms how Koide's formula is satisfied and
how the charged lepton spectrum is determined, even within EFT.
The conclusion is as follows.
Within the model to be discussed in Secs.~3--5,
the class of 1-loop diagrams shown
in Fig.~\ref{1LoopAnalyInEFT} do not generate operators that violate Koide's relation
sizably; see Sec.~5.
(There is another type of 1-loop diagrams that possibly
cancels the QED correction; see Sec.~3.)
We do not find any loop-induced higher-dimensional
operators which violate Koide's relation in conflict with
the current experimental bound.

Concerning the corrections (ii),
we will introduce specific family gauge symmetries and their breaking patterns
such that, in the first place the relation (\ref{relvi}) is satisfied at
tree level, and secondly
the corrections (ii) are suppressed.

Since the above example of underlying mechanism that suppresses
higher--dimensional operators
is simple and fairly
general, and since suppression of induced 
$1/\Lambda^n$ corrections
within EFT 
provides a non--trivial cross check of theoretical consistency,
we believe that our approach based on EFT has a certain
justification and would be useful as a basis for considering
more fundamental models.

\section{Cancellation of QED corrections}
\clfn

In this section,
we consider $U(3)$ family gauge symmetry and examine the radiative
correction to Koide's formula induced by
the family gauge interaction.
We denote the generators for the fundamental representation
of $U(3)$ by $T^\alpha$
($0\leq\alpha\leq 8$), which satisfy
\bea
{\rm tr}\left(T^\alpha T^\beta\right)=\frac{1}{2}\, \delta^{\alpha\beta} 
~~~;~~~
T^\alpha = {T^\alpha}^\dagger
\, .
\label{U3generators}
\eea
$T^0=\frac{1}{\sqrt{6}}{\bf 1}$ is the generator of $U(1)$,  
while $T^a$ ($1\leq a \leq 8$) are the generators of
$SU(3)$.
This fixes the normalization of the $U(1)$ charge.

We assign 
$\psi_L$ to the representation
$({\bf 3},1)$, where $\bf 3$ stands
for the $SU(3)$ representation and 1 for the $U(1)$ charge,
while $e_R$ is assigned to $(\bar{\bf 3},-1)$.
Under $U(3)$, the 9--component field
$\Phi$ transforms as three $({\bf 3},1)$'s.
$\varphi$ is singlet under $U(3)$.
Explicitly the transformations of these fields 
are given by
\bea
\psi_L \to U \, \psi_L \, ,
~~~
e_R \to U^* \, e_R \, ,
~~~
\Phi \to U \, \Phi \, ,
~~~
\varphi \to \varphi
\label{U3transf}
\eea
with $U = \exp \left(i\theta^\alpha T^\alpha\right)$.

We consider
\bea
{\cal O}_1=
\frac{\kappa_1(\mu)}{\Lambda^2}\,
\bar{\psi}_{L}\, \Phi\, \Phi^T\, \varphi \, e_{R} \, 
\label{exampleO1}
\eea
as the higher-dimensional operator which generates the 
charged lepton spectrum at tree level 
[corresponding to $\cal O$ of eq.~(\ref{higherdimopO})].
It is invariant under the above $U(3)$ gauge symmetry, whereas
$\cal O$ is not, since we assigned $\psi_L$ and $e_R$ to mutually conjugate
representations.
In fact, ${\cal O}_1$ is invariant under a larger symmetry  
$U(3)\times O(3)$, under which
$\Phi$ transforms as $\Phi\to U\Phi O^T$ ($O\,O^T = {\bf 1}$).
In this section we ignore the $O(3)$ symmetry and focus on the
$U(3)$ gauge symmetry.\footnote{
For definiteness, one may assume  that the $O(3)$ symmetry is gauged 
and spontaneously
broken at a high energy
scale before the breakdown of the $U(3)$ symmetry.
}
When $\Phi$ acquires a VEV\footnote{We assume that $\Phi$ can
be brought to a diagonal form by $U(3)\times O(3)$ transformation.}
\bea
\langle \Phi(\mu) \rangle
=
\left(\begin{array}{ccc}
v_1(\mu)&0&0\\
0&v_2(\mu)&0\\
0&0&v_3(\mu)
\end{array}\right) ,
\label{VEVPhiatmu}
\eea
and  if all $v_i$ are different,
$U(3)$ symmetry is completely broken by $\langle \Phi \rangle$,
and the spectrum of the $U(3)$ gauge bosons is determined 
by $v_i$.

\begin{figure}[t]\centering
\includegraphics[width=7cm]{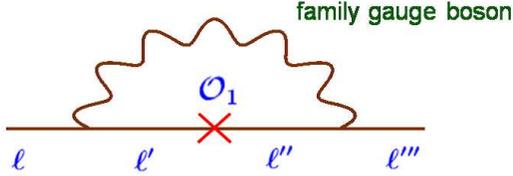}
\caption{\small 
Diagram for
the 1-loop correction to the charged lepton pole mass induced by exchange
of family gauge bosons.
\label{RadcorrByFamilyGB}}
\end{figure}

With all this setup 
we may compute the radiative corrections to the pole masses
induced by the family gauge interactions; see Fig.~\ref{RadcorrByFamilyGB}.
It turns out that the radiative corrections to the pole masses
have the same form as the QED corrections eq.~(\ref{QED1Lcorr})
with opposite sign:
\bea
\delta m^{\rm pole}_i =
-\frac{3\,\alpha_F}{8\,\pi}\left[
\log\left(
\frac{\mu^2}{v_i(\mu)^2}
\right) + c
\right] \,
{m}_i(\mu) \, ,
\label{alphaFcorr}
~~~~~~~
{m}_i(\mu) = 
\frac{\kappa_1(\mu)\,v_{\rm ew}}{\sqrt{2}\Lambda^2} \, v_i(\mu)^2 \, .
\label{mmu}
\eea
Here, $c$ is a constant independent of $i$.
$\alpha_F = g_F^2/(4\pi)$ denotes the coupling constant of 
the $U(3)$ gauge symmetry, where we assume that the
couplings of $U(1)$ and $SU(3)$ are common.\footnote{
One may worry about validity of the assumption for the universality
of the $U(1)$ and $SU(3)$ couplings,
since the two couplings are renormalized differently in general.
The universality can be ensured approximately if these two 
symmetry groups are embedded
into a simple group down to a scale close to the relevant scale.
There are more than one ways to achieve this.
A simplest way would be to embed $SU(3)\times U(1)$
into $SU(4)$ \cite{Sumino:2008hu}.
}
The Wilson coefficient
$\kappa_1(\mu)$ is defined in $\overline{\rm MS}$ scheme.
$v_i(\mu)$ are defined as follows:
The VEV of $\Phi$ at renormlization
scale $\mu$, $\langle \Phi(\mu)\rangle$
given by eq.~(\ref{VEVPhiatmu}),
is determined by minimizing the 1--loop effective potential
in Landau gauge 
(explicit form of the effective
potential will be discussed in the next section);
$\Phi$ is renormalized in $\overline{\rm MS}$ scheme.
We ignored terms suppressed by
${m}_i^2/v_j^2(\ll 1)$ in the above expression.

Some important features are as follows.
\begin{enumerate}
\item
The sign is opposite to that of
the QED correction eq.~(\ref{QED1Lcorr}),
which results from the fact that $\psi_L$ and $e_R$ have the same
QED charges but mutually
conjugate (opposite) $U(3)$ charges.
\item
Suppose the relation (\ref{relvi}) is satisfied at tree level, such that Koide's
formula is satisfied.
Then there is no ${\cal O}(\alpha_F)$ correction to this relation.
(Recall that the correction to the 1-loop effective potential
in Landau gauge \`{a} la Coleman-Weinberg is ${\cal O}(\alpha_F^2)$.)
\item
The characteristic form of the radiative corrections eq.~(\ref{alphaFcorr})
is determined by the fact that ${\cal O}_1$ is multicatively renormalized,
and also by the symmetry breaking pattern 
$U(3)\to U(2)\to U(1)\to \mbox{nothing}$.
\item
In the case that $\alpha=\frac{1}{4}\alpha_F$, the radiative 
corrections by family gauge interactions
and the QED corrections to Koide's mass formula cancel
for arbitrary $\mu$.
\end{enumerate}
Let us add some explanation on the feature 3.
\begin{figure}[t]\centering
\includegraphics[width=14cm]{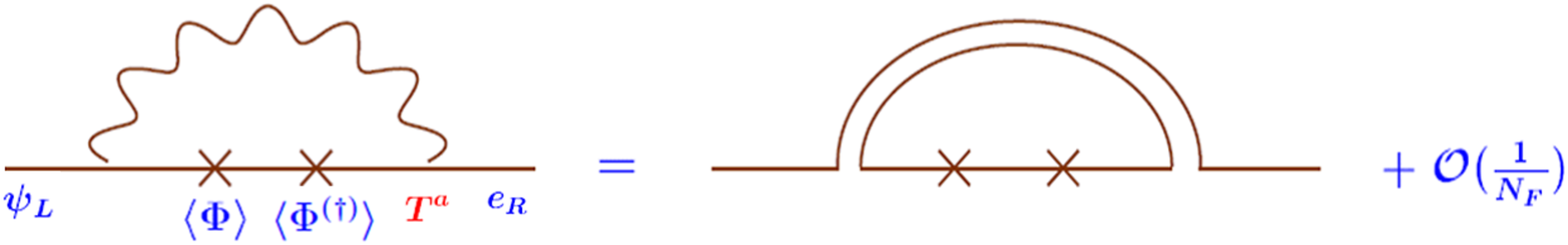}
\vspace{-2mm}\\
(a)\vspace{2mm}\\
\includegraphics[width=13cm]{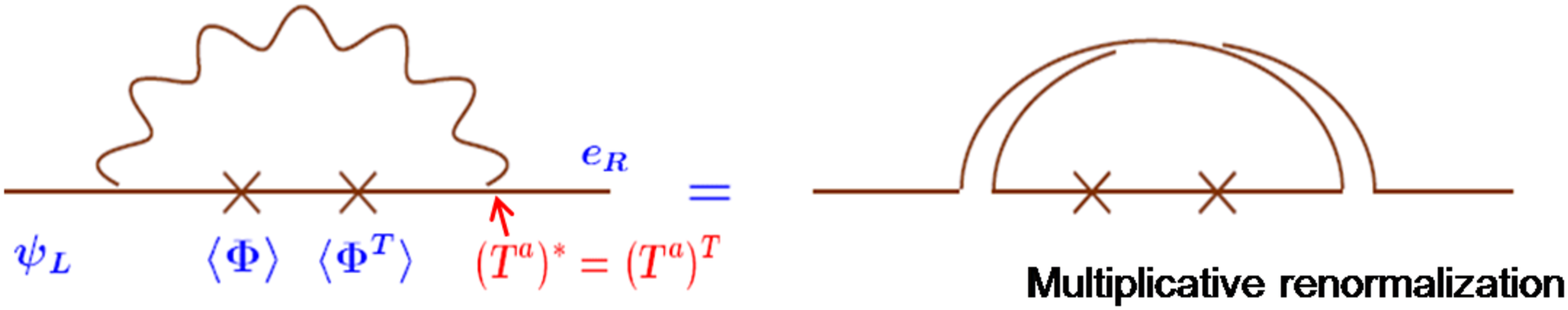}
\vspace{-2mm}\\
(b)
\caption{\small 
1--loop diagrams contributing to $\delta m_i^{\rm pole}$,
(a) in the case that
$\psi_L$ and $e_R$ are in the same representation of $SU(3)$ or $O(3)$, i.e.\
$\psi_L:({\bf 3},Q)$ and $e_R:({\bf 3},Q')$, and
(b) in the case that $\psi_L$ and $e_R$ are in the conjugate representations
of $U(3)$, i.e.\
$\psi_L:({\bf 3},1)$ and $e_R:(\bar{\bf 3},-1)$.
The right--hand--sides show flows of family charge.
In (a), the closed loop of the family charge flow
represents ${\rm tr}\bigl(\Phi\Phi^{(\dagger)}\bigr)$.
In (b), the charge flow is connected in one line, 
which has the same form as the tree diagram,
showing multiplicative
renormalization of the operator ${\cal O}_1$.
\label{DoubleLineDiagram}}
\end{figure}
The operator ${\cal O}_1$ is the only dimension-6 operator invariant
under $U(3)\times O(3)$, so it should be renormalized multiplicatively.
In this regard, a pedagogical comparison 
is shown in Figs.~\ref{DoubleLineDiagram}(a)(b).
Had we chosen the same representation for $\psi_L$ and $e_R$ under
family symmetry such as $SU(3)$ or $O(3)$, the dimension-4 operator
$\bar{\psi}_{Li} \varphi e_{Ri}$ would be allowed  by symmetry
\cite{Antusch:2007re}.
In fact, the 1--loop diagram shown in Fig.~\ref{DoubleLineDiagram}(a) induces an effective 
operator
\bea
{\cal O} '\sim \frac{\alpha_F}{\pi}\times{\kappa}\,
\bar{\psi}_{Li} \, \varphi \, e_{Ri} 
\times \frac{\langle\Phi\rangle_{jk}
\langle\Phi^{(\dagger)}\rangle_{kj}
}{\Lambda^2} 
\, ,
\eea
hence corrections universal to all the charged--lepton masses,
$(\delta m_e, \delta m_\mu, \delta m_\tau) \propto (1,1,1)$,
are induced.
This correction changes the direction of 
$(\sqrt{m_e},\sqrt{m_\mu},\sqrt{m_\tau})$ and
violates Koide's formula rather strongly;
moreover the correction 
is dependent on the cut-off $\Lambda$ of the loop integral.
In order that the correction to Koide's formula cancel the QED correction,
a naive estimate shows that 
$\alpha_F/\pi$ should be order $10^{-5}$, provided that
$\Lambda$  is not too large.
By contrast, 
in the case that
$\psi_L$ and $e_R$ are assigned to the conjugate representations
of $U(3)$, the charge flow is connected in one line, so that it has the same
charge flow structure as the tree graph.
The form $\sim \log\mu \times m_i$ of the 1-loop correction
can be understood in this way.

$v_i^2$ in the argument of log in eq.~(\ref{alphaFcorr}) stems from IR
cut-off of the loop integral.
Namely they come from the masses of family gauge bosons.
As $v_3>v_2>v_1>0$ are successively turned on, family symmetry breaks according
to the pattern
$U(3)\to U(2)\to U(1)\to \mbox{nothing}$.
Gauge bosons corresponding to broken generators decouple
at each stage, and their masses enter the argument of log as IR cut-off.
The form of eq.~(\ref{alphaFcorr}) is essentially determined by this
symmetry breaking pattern. (See \cite{Sumino:2008hy} for a more precise argument.)
The same symmetry breaking pattern resides in the QED Lagrangian: as
$m_\tau > m_\mu >m_e>0$ are successively turned on,
chiral symmetry breaks according to $U(3)\to U(2)\to U(1)\to \mbox{nothing}$.
Essentially this is the reason why the two radiative corrections have the same form.

\begin{figure}[t]\centering
\begin{tabular}{ccc}
\includegraphics[width=7cm]{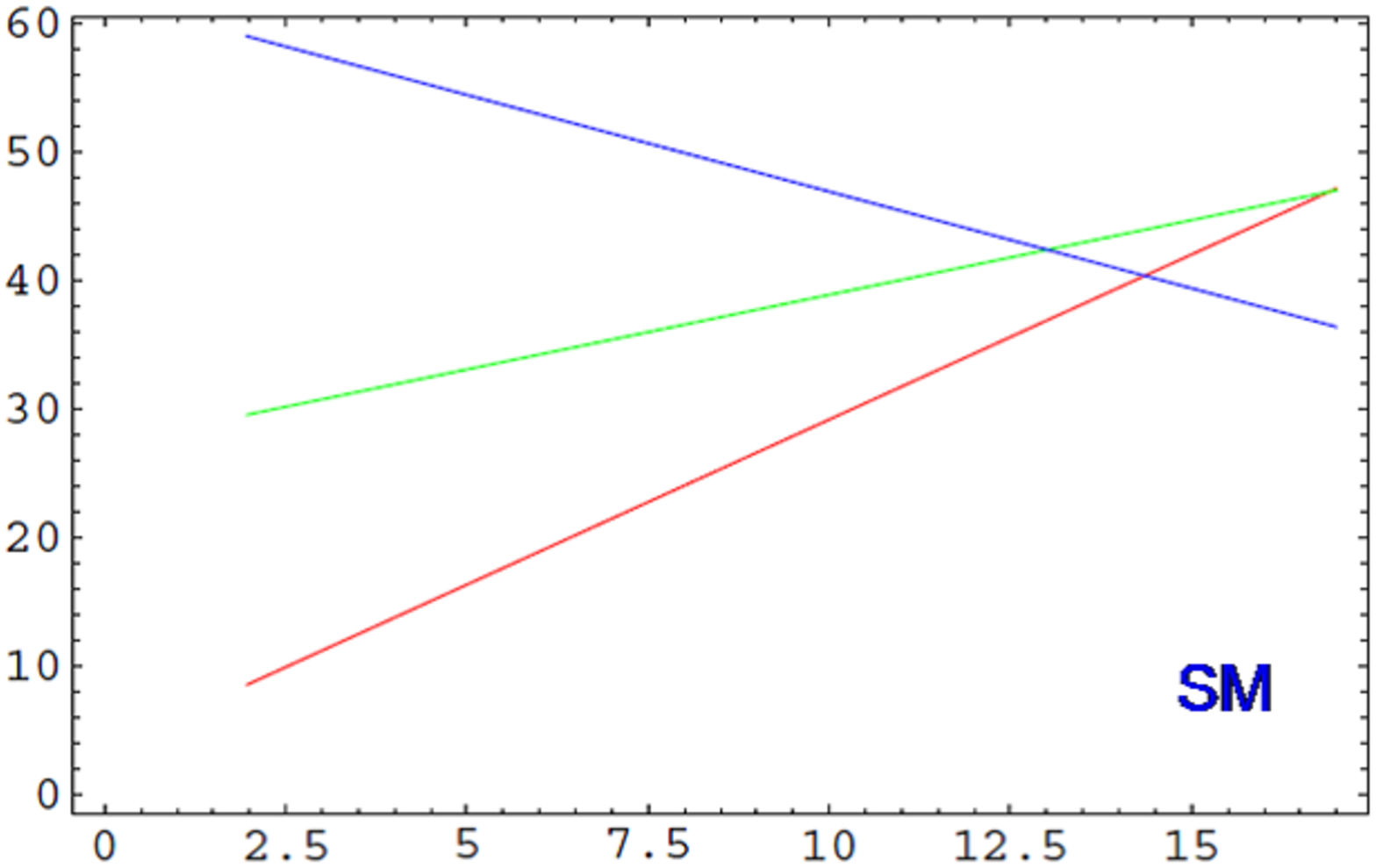}&~~~&
\includegraphics[width=7cm]{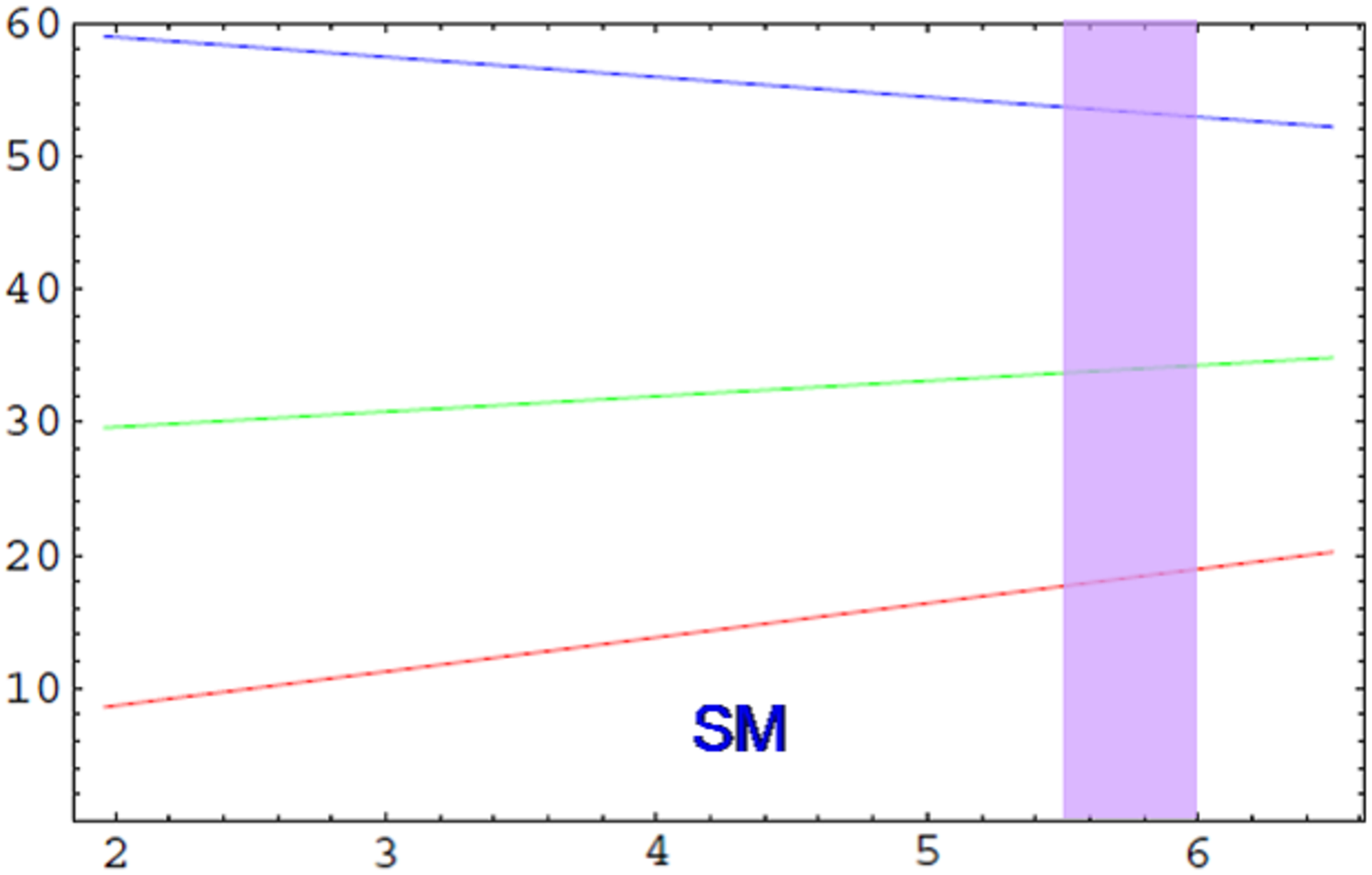}\\
(a)&&(b)
\end{tabular}
\caption{\small 
Inverse of the three gauge couplings of the SM 
$\alpha_1^{-1},\alpha_2^{-1},\alpha_3^{-1}$ vs.\
$\log_{10}(\mu/{\rm GeV})$:
(a) from electroweak scale to GUT scale, and (b) from electroweak scale up to
$10^3$~TeV.
The green line corresponds to the relevant coupling $\alpha_W$.
The shaded
band shows an allowed variation range of the unification scale
in order to meet the present experimental accuracy of Koide's formula.
\label{RunningCouplings}}
\end{figure}

Now we speculate on a possible scenario how the relation (feature 4)
\bea
\alpha\approx\frac{1}{4}\alpha_F
\label{relalphas}
\eea
 may be satisfied.
In fact, this relation should be satisfied within 1\% accuracy,
in order that Koide's relation is satisfied within the present
experimental bound.
As already discussed, the scale of $\alpha$ is determined by the charged
lepton masses, while the scale of $\alpha_F$ is determined by 
the family gauge boson masses, which should be much
higher than the electroweak 
scale.
Since the relevant scales of the two couplings
are very different, we are unable to 
avoid assuming some
accidental factor (or parameter tuning) to achieve this condition.
Instead we seek for an indirect evidence which indicates such an
accident has occurred in Nature.
The relation (\ref{relalphas}) shows that the value of
$\alpha_F$ is close to that of the
weak gauge coupling constant $\alpha_W$, since $\sin^2\theta_W(M_W)$ is close
to $1/4$.
In fact, within the SM, $\frac{1}{4}\,\alpha_W(\mu)$ approximates $\alpha(m_\tau)$
at scale $\mu \sim 10^2$--$10^3$~TeV.
Hence, if the electroweak $SU(2)_L$ gauge
group and the $U(3)$ family gauge
group are unified 
around this scale, naively we expect that
$
\alpha \approx \frac{1}{4}\, \alpha_F
$
is satisfied.
Since $\alpha_W$ runs relatively slowly in the SM, 
even if the unification scale is varied within a factor of 3,
Koide's mass formula is satisfied within the present experimental
accuracy.
This shows the level of parameter tuning required in this scenario.
Fig.~\ref{RunningCouplings}(a)(b) show the running of the inverse of the
three gauge couplings of the SM.
The former figure is well-known, which shows the running 
from the electroweak scale up to GUT scale.
The latter figure shows the same running up to $10^3$~TeV.
The shaded band shows an allowed variation range (factor 3)
of the unification scale, which is
limited by the present experimental accuracy of Koide's formula.

\section{Potential predicting Koide's formula and realistic lepton spectrum}
\clfn

In this section we study how the relation (\ref{relvi}) can be satisfied by
the VEV of $\Phi$.
For this purpose, we
introduce a model of charged lepton sector based on
$U(3)\times O(3)$ family gauge symmetry,
within EFT valid at scales below $\Lambda$.
The choice of this family symmetry is motivated by the fact that 
this is the largest symmetry
possessed by the operator ${\cal O}_1$ analyzed in the
previous section.
In particular, in this section we focus on the potential of scalar
fields.
In the first place,
we find that the $U(3)\times O(3)$ family symmetry
is not sufficiently restrictive.
Namely, since this symmetry does not constrain
the potential of 
$\Phi$ sufficiently,
one needs to tune the parameters of the potential in order to realize the
relation (\ref{relvi}).
We need some symmetry enhancement in order to realize this relation
without fine tuning.

\begin{figure}[t]\centering
\psfrag{45}{\hspace{0mm}$45^\circ$}
\psfrag{Phi0}{\hspace{0mm}$\Phi^0T^0$}
\psfrag{Phia}{\hspace{0mm}$\Phi^aT^a$}
\psfrag{Phimu}{\hspace{0mm}$\Phi^\alpha T^\alpha$}
\includegraphics[width=6cm]{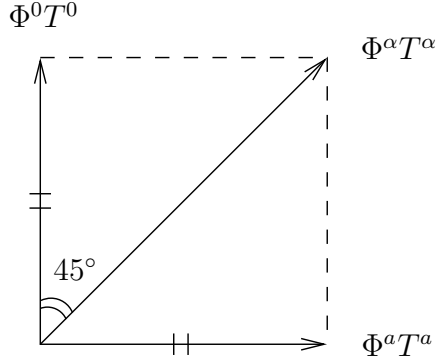}
\caption{\small 
Geometrical interpretation of \protect{eq.~(\ref{KFrelation})}.
\protect{Eq.~(\ref{U3generators})} defines the inner product in a 9--dimensional
real vector space spanned by the basis $\{ T^\alpha \}$.
Since $\Phi^0 T^0$, $\Phi^a T^a$ and $\Phi=\Phi^\alpha T^\alpha$
form an isosceles right triangle,
the angle between $T^0$ and $\Phi$ is $45^\circ$.
This is
Koide's formula in the basis where $\Phi$ is diagonal.
\label{GeometricInt}}
\end{figure}
In order to find an appropriate larger symmetry, we recall the conditions
equivalent to the relation (\ref{relvi}).
Let us express components of  $\Phi$ using $T^\alpha$, introduced in 
eq.~(\ref{U3generators}),
as the basis:
\bea
\Phi =  \Phi^\alpha \, T^\alpha \, .
\eea
In general $\Phi^\alpha$ take complex values.
As shown by Koide \cite{Koide:1989jq},
if 
\bea
(\Phi^0)^2= \Phi^a\, \Phi^a
~~~~~;~~~~~
\Phi^\alpha \in {\bf R}\, 
\label{KFrelation}
\eea
are satisfied,
the relation (\ref{relvi}) is satisfied
by the eigenvalues of $\Phi$.
There is a geometrical interpretation of these conditions in terms
of a real vector in a 9-dimensional space; 
see Fig.~\ref{GeometricInt}.
This picture indicates that a symmetry associated with this
9-dimensional space may be relevant.

\begin{figure}[t]\centering
\includegraphics[width=8cm]{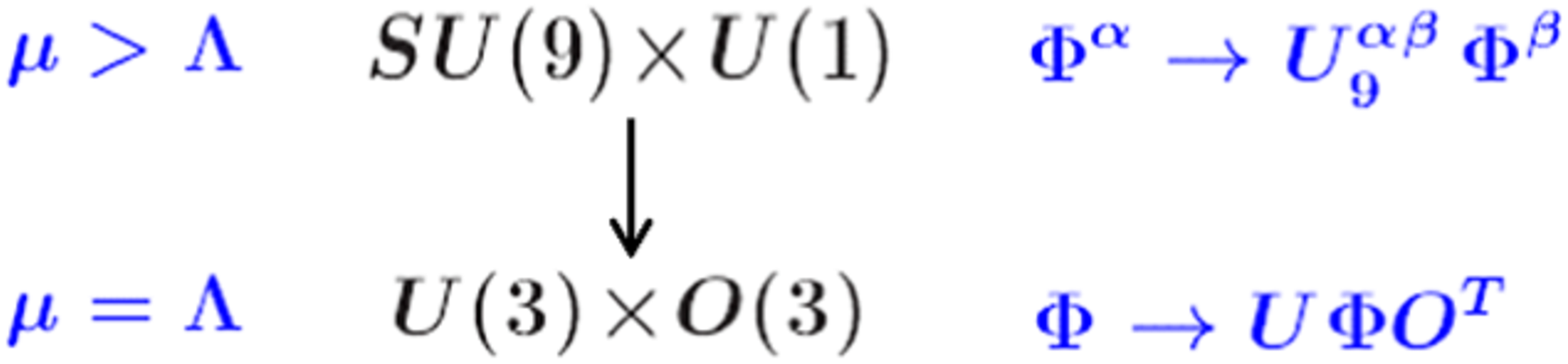}
\caption{\small 
Symmetry breaking pattern assumed in our senario.
\label{SymEnhanceSenario}}
\end{figure}

Motivated by this picture, we adopt 
$SU(9)\times U(1)$ as an enhanced symmetry, where
the $SU(9)\times U(1)$ transformation is given by
$\Phi^\alpha \to U_9^{\alpha\beta}\,\Phi^\beta$
($U_9$ is a 9-by-9 unitary matrix).
Then we assume the following symmetry breaking scenario:
Above the cut--off scale $\Lambda$ there is an $SU(9)\times U(1)$ 
gauge symmetry, and this
symmetry is spontaneously broken to
$U(3)\times O(3)$ below the cut--off scale; see Fig.~\ref{SymEnhanceSenario}.
One can check that indeed 
$U(3)\times O(3)$ is a subgroup of $SU(9)\times U(1)$.
In what follows,
we do not discuss any specific model at scales
above $\Lambda$ but only assume restoration of this
larger symmetry.

Within this senario, we still need to
introduce an additional scalar field
$X$
in order to realize a desirable vacuum configuration.
Under $SU(9)\times U(1)$, $X$ is in the representation
$({\bf 45},Q_X)$ (the ${\bf 45}$ is the
second--rank symmetric representation) and is unitary.
It can be represented by a 9--by--9 unitary symmetric matrix:
\bea
X^{\alpha\beta}=X^{\beta\alpha},
~~~~~
X^{\alpha\gamma}\,{X^{\beta\gamma}}^* = \delta^{\alpha\beta} 
~~~~~;~~~~~
X^{\alpha\beta}\to
{U}_9^{\alpha\rho}\,X^{\rho\sigma}\,{U}_9^{\beta\sigma} \, .
\label{transfX}
\eea
Note that the unitarity condition is compatible with the symmetry 
transformation.

We have analyzed 
the potential of $\Phi$ and $X$ that is invariant
under $U(3)\times O(3)$ and consistent with the assumed symmetry
breaking pattern shown schematically in Fig.~\ref{SymEnhanceSenario}.
The upshot is that in a finite region of the parameter space of the
potential Koide's relation is satisfied by the eigenvalues of 
$\langle \Phi \rangle$.
Furthermore, the eigenvalues can be made consistent with 
the experimental values of the charged lepton masses without 
fine tuning of parameters.
In the rest of this section we briefly sketch the argument.
(See \cite{Sumino:2008hy} for details.)

Operators in
the potential which are
invariant under $SU(9)\times U(1)$ read
\bea
&&
\tilde{V}_\Phi = -m^2\,{\Phi^\alpha}^*\Phi^\alpha
+\lambda\,
({\Phi^\alpha}^*\Phi^\alpha)^2 + \cdots\, ,
\\ && 
\tilde{V}_X = {\rm const.}\, ,
\\ &&
\tilde{V}_{\Phi X}=
\varepsilon_K \, \bigl| \Phi^\beta \, {X^{\beta\gamma}}^* \, \Phi^\gamma \bigr|^{2}
+ \cdots \, ,
\eea
where only some representative terms have been shown explicitly.
In particular, we take into account operators with dimensions higher than 4, 
although they
are not shown explicitly.
That $\tilde{V}_X={\rm const.}$ follows from the fact that $X$ is unitary
and has a non-vanishing $U(1)$ charge.
Similarly operators invariant under $U(3)\times O(3)$ but non-invariant
under $SU(9)\times U(1)$ read
\bea
 &&
V_\Phi = g_1\,{\rm tr}(\Phi^\dagger\,\Phi\,\Phi^\dagger\,\Phi) +g_2\,
{\rm tr}(\Phi\,\Phi^T\,\Phi^*\,\Phi^\dagger)
 + \cdots \, ,
\\  &&
V_X =  h_1\,{\rm
tr}(T^\alpha\,T^\rho\,T^\beta\,T^\sigma)\,
X^{\alpha\beta}\,{X^{\rho\sigma}}^* \, + \cdots\, ,
\\  &&
V_{\Phi X}= \cdots \, .
\eea

After examination, we find that, in a certain parameter region,
the global minimum of $V_X$ is located at the configuration\footnote{
To be more accurate, one needs to impose
either approximate or (spontaneously-broken)
exact $CP$ symmetry in addition to $U(3)\times O(3)$ symmetry.
}
\bea
\langle X^{\alpha\beta}\rangle = {\rm diag.}
(-1,\underbrace{+1,\cdots,+1}_{{8}}) 
=-2\,\delta^{\alpha 0}\delta^{\beta 0}+\delta^{\alpha\beta}\, ,
\eea
and
$\tilde{V}_{\Phi X}$ is minimized in the case
\bea
\Phi^\beta \, {X^{\beta\gamma}}^* \, \Phi^\gamma=0\, ,
\eea
as may be inferred from the term of $\tilde{V}_{\Phi X}$ shown explicitly.
One observes that
if the above two equations are combined,
Koide's relation
\bea
(\Phi^0)^2= \Phi^a\, \Phi^a
\eea
follows.
(Reality of $\Phi$ can also be derived from $\tilde{V}_{\Phi}$ and $V_{\Phi X}$,
but we skip explanation of this part.)

Let us first give an argument which is not very solid but more illustrative.
The above observation shows that if there is a hierarchy
of parameters given by
\bea
\tilde{m}^2,\lambda,\varepsilon_K, \tilde{h}_1 \gg g_1, g_2 \, ,
\label{HierarchyOfParam}
\eea
while ignoring all the other parameters,
Koide's mass formula will be satisfied approximately.
Here, 
$\tilde{m}^2$ and $\tilde{h}_1$ denote dimensionless couplings
defined from $m^2$ and $h_1$, respectively,
after rescaling them by an appropriate mass scale (e.g.\ $v_3$).
Furthermore, we find that
if there is an additional hierarchy given by
\bea
\tilde{m}^2,\lambda,\varepsilon_K, \tilde{h}_1 \gg g_2\gg g_1 \, ,
\label{HierarchyOfParam2}
\eea
a realistic charged lepton spectrum
$v_1:v_2:v_3 = \sqrt{m_e}:\sqrt{m_\mu}:\sqrt{m_\tau}$
can be explained without fine tuning.
(This follows from an explicit computation, and we do not know 
any other simple
explanation.)
Naively one may
expect that the parameters in the $SU(9)\times U(1)$ non-invariant
operators $g_1,g_2,\tilde{h}_1$
 are suppressed as compared to the parameters in the
$SU(9)\times U(1)$ invariant operators 
$\tilde{m}^2,\lambda,\varepsilon_K$, since
the former parameters are generated only
through spontaneous symmetry breaking
at the cut-off scale.
But this is not necessarily the case for parameters such as
$\tilde{h}_1$ that are originally
dimensionful parameters.
Let us speculate on a possible underlying mechanism
that would lead to a hierarchy of potential parameters
given by eq.~(\ref{HierarchyOfParam2}).

Suppose that the symmetry breaking
$SU(9)\times U(1)\to U(3)\times O(3)$
is induced by a condensate of a scalar field
$T^{\alpha\beta}_{\rho\sigma}$, which is
a 4th-rank tensor 
under $SU(9)$.
Indeed if 
$\langle T^{\alpha\beta}_{\rho\sigma} \rangle \sim
{\rm tr}(T^\alpha{T^\beta}^*{T^\rho}^*T^\sigma)$,
this symmetry breaking takes place.\footnote{
By analyzing the potential of $T^{\alpha\beta}_{\rho\sigma}$
up to quartic terms,
we have checked that in a certain parameter region of the potential,
$\langle T^{\alpha\beta}_{\rho\sigma} \rangle \sim
{\rm tr}(T^\alpha{T^\beta}^*{T^\rho}^*T^\sigma)$ becomes
a local minimum of the potential.
(We were unable to clarify whether this can be a global minimum,
due to technical complexity.)
}
\begin{figure}[t]\centering
\includegraphics[width=15cm]{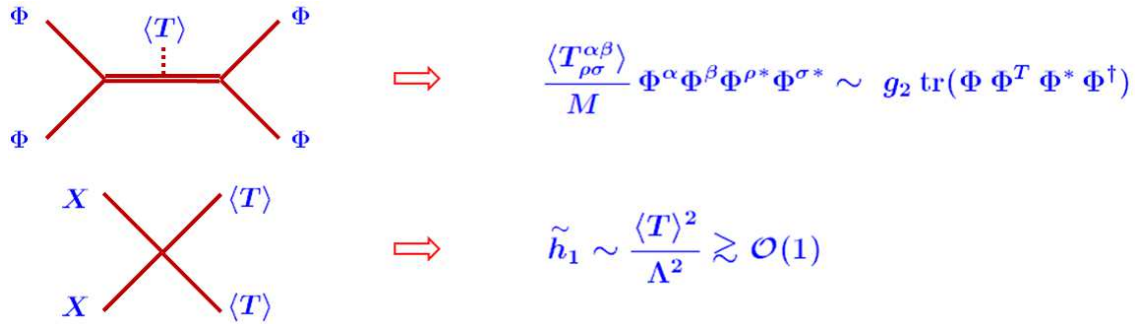}
\caption{\small 
Speculation on underlying physics
that may generate $SU(9)\times U(1)$ non-invariant
operators.
\label{SpeculationGenPot}}
\end{figure}
Through the first diagram shown in Fig.~\ref{SpeculationGenPot},
the operator $g_2\,
{\rm tr}(\Phi\,\Phi^T\,\Phi^*\,\Phi^\dagger)$
may be induced; the double line denotes a heavy degree of freedom
with an $SU(9)\times U(1)$-invariant mass scale $M$.
Since $\langle T\rangle \sim {\cal O}(\Lambda)$, the coefficient
$g_2\sim \Lambda/M$ would be a small parameter provided $M\gg \Lambda$.
$g_1$ is even more suppressed, since the operator 
$g_1\,{\rm tr}(\Phi^\dagger\,\Phi\,\Phi^\dagger\,\Phi)$
cannot be generated by a single insertion of $\langle T\rangle$
at tree-level.
Either two insertions of $\langle T\rangle$ or a loop correction
is necessary,
which leads to additional suppression factors.
The second diagram in Fig.~\ref{SpeculationGenPot} would induce
the operator 
$h_1\,{\rm
tr}(T^\alpha\,T^\rho\,T^\beta\,T^\sigma)\,
X^{\alpha\beta}\,{X^{\rho\sigma}}^*$
(together with other operators).
Since there is no intermediate heavy degree of freedom,
the induced coupling $h_1$, when normalized by $\Lambda$,
would be order 1.

We turn to a more solid argument.
It is legitimate to identify
the above potential to be the 
effective potential (including loop corrections and in Landau gauge).
Recall that, according to the argument in Sec.~3, 
we can connect the pole masses of the charged leptons to
the VEV of $\Phi$, which is determined from the 
effective potential renormalized at an
arbitrary scale $\mu$.\footnote{
This is a consequence of the fact that $\Phi$ is renormalized multiplicatively,
namely the counter terms for all $v_i$ are common.
More precisely, physically it is adequate to choose
the scale $\mu$ to be larger than the family gauge boson
masses in $\overline{\rm MS}$ scheme.
}
For our purpose,
it is most convenient to set the scale to be $\mu=\Lambda$, since
at this scale radiative corrections within EFT essentially vanish,
and the parameters of the effective potential are set
by the boundary (initial) conditions derived from the theory
above the cut-off scale.

The following conclusions have been drawn from a detailed analysis of
the general potential of $\Phi$ and $X$.
In the case that 
certain hierarchical relations among the parameters of the potential are satisfied,
Koide's formula as well as a realistic charged lepton spectrum follow,
consistently with the present experimental values.
These hierarchical 
conditions on the parameters
are a generalization of eq.~(\ref{HierarchyOfParam2}).
Typical sizes of the required hierarchies of the potential parameters
are of order $10^{-3}$--$10^{-4}$.
These hierarchical relations are consistent with the assumed symmetry and
symmetry enhancement.
Namely, those parameters which need to be suppressed are associated
with $SU(9)\times U(1)$ non-invariant operators.
Their values at the boundary
$\mu=\Lambda$ are determined by the dynamics
above the cut-off scale.
On the other hand, 
up to now
there exists no model of the scales above the cut-off, $\mu>\Lambda$, 
which leads to these
hierarchical relations among the potential parameters.
(Nothing more than the speculation given 
in Fig.~\ref{SpeculationGenPot} exists.)

Finally we comment on how a realistic 
charged lepton spectrum follows without fine tuning of parameters.
Koide's formula imposes one relation among the three
charged lepton masses.
Hence, apart from an overall dimensionful scale of the three masses, 
there remains
one-parameter degree of freedom.
Suppose this degree of freedom is fixed 
by minimizing the operator
$V_{\Phi 3}=g_2\,
{\rm tr}(\Phi\,\Phi^T\,\Phi^*\,\Phi^\dagger)$.
Then, $m_e/m_\tau$ is predicted to be 15\% away from the experimental 
value, and $m_\mu/m_\tau$ is 
predicted to be 1.5\% away from the experimental 
value.
In other words, already the values are close to the true values
and orders of magnitude of the mass ratios are
predicted correctly.
When any other $SU(9)\times U(1)$ non-invariant
operators, which modify the potential minimum,
are turned on, as long as contributions of these operators
are suppressed as compared to $V_{\Phi 3}$,
the values of $m_e/m_\tau,m_\mu/m_\tau$ do not alter significantly.
Since Koide's relation is protected by the large 
couplings (e.g.\ $\tilde{m}^2,\lambda,\varepsilon_K, \tilde{h}_1$),
and since Koide's relation is satisfied experimentally,
with some small 
values of parameters,
$m_e/m_\tau$ and $m_\mu/m_\tau$ are designated to
coincide with the experimental values.
This is not regarded as fine tuning.
The overall scale of the lepton masses is determined by
the parameters of $\tilde{V}_{\Phi}$, such as $m^2$ and $\lambda$, and
by $\kappa_1$ of the operator ${\cal O}_1$.
Since $v_3/\Lambda\simgt 0.1$ is not extremely small,
there is no fine tuning problem within EFT for predicting the
overall scale.

\section{Summary and discussion}
\clfn

Let us summarize our study.
We have analyzed the role of family gauge symmetries in
relation with the charged lepton spectrum and Koide's mass formula.
The analysis is performed within EFT valid below a cut-off scale $\Lambda$
and within the known scenario in which the charged lepton mass matrix
is proportional to $\langle \Phi \rangle^2$ at leading order.
Before describing the analysis,
we made an argument to justify usefulness of an EFT approach
and to discuss why EFT does not instantly run into problems by
higher order corrections in $1/\Lambda$.

In the first part of our analysis,
we studied radiative corrections generated by 
$U(3)$ family gauge interaction.
$U(3)$ family gauge symmetry has a unique property
with respect to the radiative correction to Koide's formula.
In fact, if $\psi_L$ and $e_R$ are assigned to 
mutually conjugate representations,
the $U(3)$ radiative correction has the same form as the QED
correction with opposite sign.
In particular, if $\alpha=\frac{1}{4}\alpha_F$,
both corrections cancel.
We discussed this possibility within a scenario in which 
$U(3)$ family symmetry is unified with $SU(2)_L$ symmetry
at $10^2$--$10^3$~TeV scale.

Some key aspects which led to the non-trivial form of the
radiative corrections
are as follows.
(1) Multiplicative renormalizability of the operator
${\cal O}_1=
\frac{\kappa}{\Lambda^2}\,
\bar{\psi}_{L}\, \Phi\, \Phi^T\, \varphi \, e_{R}$
ensures that only logarithmic corrections to the lepton mass matrix
appear;
furthermore, multiplicative renormalizability of 
$\langle \Phi \rangle$ ensures that the correction to
Koide's formula is
independent of the renormalization scale $\mu$ of
the effective potential.
(2) The symmetry breaking pattern
$U(3) \to U(2)\to U(1)\to \mbox{(nothing)}\,$ essentially
dictates how the IR cut-off enters the logarithmic correction
at each stage of the symmetry breaking;
this symmetry breaking pattern happens to be common
to the family gauge sector and QED sector.
(3) We assumed that $\langle \Phi \rangle$ can be brought to 
a diagonal form
by symmetry transformation and also
the tree-level Koide's relation for
the diagonal elements:
$
\frac{v_1(\mu)+v_2(\mu)+v_3(\mu)}
{\sqrt{3\,[v_1(\mu)^2+v_2(\mu)^2+v_3(\mu)^2]}}
=\frac{1}{\sqrt{2}}
$.

In the latter part
of the analysis, we have examined how this relation 
among $v_i$ may be realized.
We proposed a potential model within EFT with a family symmetry
$U(3)\times O(3)$.
Motivated by a geometrical interpretation of Koide's relation
(c.f.\ Fig.~\ref{GeometricInt}),
we further imposed symmetry enhancement to $SU(9)\times U(1)$
above the cut-off scale.
We have introduced another scalar field $X$ and examined
the general potential of $\Phi$ and $X$.
In this manner, we were able to find a potential minimum
which leads to Koide's formula and
a realistic charged lepton spectrum.
The potential parameters need to satisfy certain
hierarchical relations
at the boundary $\mu=\Lambda$ of EFT, which are consistent with
the symmetry requirements.
We have speculated on underlying
physics which may lead to
the hierarchical relations.
\medbreak

There are many unsolved questions in the present analysis.
The list is as follows.
\begin{itemize}
\item
Quarks and neutrinos are not included in the analysis.
In relation to this, with the fermion content discussed in this
analysis, anomalies induced by the family gauge interactions
do not cancel.
\item
$O(3)$ gauge symmetry needs to be broken spontaneously above the 
$U(3)$ symmetry breaking scale, in order to suppress mixing of
gauge bosons of both gauge groups.
We have not implemented a mechanism to achieve this.
\item
The VEV
$\langle \Phi\rangle$, which explains the realistic charged lepton
spectrum, cannot be diagonalized by $U(3)\times O(3)$ symmetry
transformation.
In order to realize the scenario for the cancellation of QED correction,
we need, for instance, to introduce another field $\Sigma$ and
its potential with $\Phi$ \cite{Sumino:2008hy}
and replace the operator ${\cal O}_1$ as
$
\frac{\kappa}{\Lambda^2}\,
\bar{\psi}_{L}\, \Phi\, \Phi^T\, \varphi \, e_{R} \to
\frac{\kappa}{\Lambda^3}\,
\bar{\psi}_{L}\, \Phi\, \Sigma\,\Phi^T\, \varphi \, e_{R} 
$.
\item
In the first part of our analysis, we considered unification of 
$U(3)$ family symmetry and $SU(2)_L$ weak symmetry.
In the latter part, we assumed embedding $U(3)\times O(3)$ into 
$SU(9)\times U(1)$.
How to make both scenario compatible has not been addressed.
It would require a large symmetry group above the cut-off scale.
\item
Fine tuning is required to stabilize small scales compared to
the cut-off scale of EFT $\Lambda\sim 10^3$~TeV.
These small scales are the VEVs $\langle \varphi \rangle$, (physical scale\footnote{
Since we normalized $X$ to be dimensionless
in eq.~(\ref{transfX}), physical scale of 
$X$ is determined by the normalization of the kinetic term
$f_X^2 \partial_\mu {X^{\alpha\beta}}^* \partial_\mu X^{\alpha\beta}$.
In order that the spectrum of $U(3)$ gauge bosons be determined mostly by
$\langle \Phi \rangle$, a hierarchy
$f_X,\langle \Sigma \rangle \ll v_1$ is required.
}
of) $\langle X \rangle$ and $\langle \Sigma \rangle$.
This fine tuning problem is similar to that of the SM.
\item
Models above the cut-off scale are completely missing.
\end{itemize}
There seems to be certain solution(s) to each of these problems
(except for the last two problems), if
we extend our model in a sufficiently complicated manner.
On the other hand, it seems very difficult to solve all of them
in a simple and unified way.

We have made a non-trivial consistency check of our present analysis.
Using the potential of $\Phi$ and $X$, we have identified
the mass eigenstates of scalar fields at the vacuum;
then we have
computed 1-loop radiative correction to the operator 
${\cal O}'_1=\frac{\kappa}{\Lambda^3}\,
\bar{\psi}_{L}\, \Phi\, \Sigma\,\Phi^T\, \varphi \, e_{R} 
$ by these scalars.
This corresponds to incorporating the class of diagrams shown in
Fig.~\ref{1LoopAnalyInEFT}.
The correction to Koide's formula turned out to be quite
suppressed and does not conflict the present experimental bound.
\medbreak

In view of the many problems listed above,
it seems quite unlikely that our model, as a whole, correctly describes
the true mechanisms of generation of the charged lepton spectrum.
Nevertheless, we suspect that
some of the mechanisms which we proposed may
reflect the true aspects of Nature.
We consider the following feature particularly non-trivial:
Not only Koide's formula but also $m_e/m_\tau$ and $m_\mu/m_\tau$
can be explained without fine tuning.
Since Koide's relation treats $m_e,m_\mu,m_\tau$
symmetrically, in many models hierarchical structure
of the spectrum is difficult to realize compatibly with Koide's
relation, without fine tuning of parameters.
As a final remark, we note that some phenomenological implications
of the present scenario have been discussed in \cite{Sumino:2008hy}.

\section*{Acknowledgements}

The author thanks K.~Tobe for discussion.
The author is also grateful to K.~Fujii for his hospitality
at KEK while part of this work has been completed and 
for listening to the argument patiently.

\end{document}